\documentclass[10pt]{iopart}
 \usepackage{iopams,graphicx}
 \begin{document}
 \newcommand{\bra}{\left \langle}
 \newcommand{\ket}{\right \rangle}
 \newcommand{\Green}[1]{\bra r \left | G_{\ell}^{#1}(k) \right | r'\ket}
 \newcommand{\Ref}[1]{\mbox{(\ref{#1})}}
 \newcommand{\Half}{\textstyle{\frac{1}{2}}}
 \title{ Effective range function below threshold}
 \author{A Deloff
 \footnote[1]{email: deloff@fuw.edu.pl}}
 \address{
 Soltan Institute for Nuclear Studies, Hoza 69, 00-681 Warsaw, Poland}
 \begin{abstract}
 We demonstrate that
 the kernel of the Lippmann-Schwinger equation, associated with
 interactions consisting of a sum of the Coulomb plus a short range
 nuclear potential, below threshold becomes degenerate. Taking advantage
 of this fact, we present a simple method of calculating the effective
 range function for {\em negative} energies. This may be useful in
 practice since the effective range expansion extrapolated to threshold
 allows to extract the low-energy scattering parameters: the Coulomb
 modified scattering length and the effective range.
 \end{abstract}
 \pacs{03.65.Ge, 02.03.Rz, 21.10.Sf, 36.10.Gv}
 \maketitle
 \par
  \section{Introduction}
  Effective range function plays  central role in the analysis and
  interpretation of two-particle low-energy scattering data. It is well
  known from the simplest case of s-wave scattering by a short range
  potential that it is more advantageous to study the effective range
  function $k\,\cot{\delta}$ than the phase shift $\delta$.
  The former is an even function of the momentum $k$ and close
  to threshold can be expanded in a power series of $k^{2}$, known as 
  the effective range expansion. The first two coefficients in this
  expansion: the scattering length and the effective range,
  respectively, carrying important information about the underlying
  interaction, are regarded as fundamental parameters in low-energy
  scattering phenomenology. The inclusion of the long range
  Coulomb interaction is a non-trivial complication because the latter
  changes dramatically the singularity structure of the scattering
  matrix. The Coulomb corrections to the effective range parameters
  are not only model dependent but can be also quite sizable. Especially
  sensitive is the scattering length where, strictly speaking, there is
  even no upper limit to the size of the Coulomb correction, i.e. 
  the scattering length that is finite in absence of Coulomb interaction,
  could become infinite when Coulomb interaction is turned on.
   Perhaps the best illustration that in real life Coulomb
   correction can be quite large is provided by
  the nucleon-nucleon scattering. 
  With exact isospin invariance the nn and pp $\mbox{}^{1}S_{0}$
  scattering lengths  in absence of Coulomb interaction should be equal
  but experimentally \cite{E1,E2,E3}: $a_{nn}=-18.6\pm 0.5\;$fm and 
  $a_{pp}=-7.828\pm 0.008\;$fm. It has been confirmed by model
  calculations that most of the difference in these two values
  may be attributed to the Coulomb interaction.
  \par
  The analytic continuation of the effective range function to negative
  energies presents considerable interest as it provides simple means
  to locate the near-threshold singularities of the scattering matrix,
  that might be associated with shallow bound states. For conducting a direct
  calculation of the effective range function below threshold, the
  standard Coulomb wave functions for negative energies are required, and,
  efficient algorithms for calculating 
  them are currently available in the literature \cite{Ian1,Ian2}.
  Nevertheless, this approach
  leads to difficulties as the standard Coulomb wave functions are
  ill chosen for the purpose and for attractive Coulomb potential
  they are singular at the Coulomb bound state energies. Therefore,
  the usual procedure employs the effective range expansion for
  effecting the extrapolation to negative energies.
  \par
  In this work we  wish to present a direct method of calculating the
  effective range function below threshold based on the solution
  of the Lippmann-Schwinger equation. In this equation the Coulomb
  potential does not appear explicitly but has been accounted for
  exactly by introducing Coulomb-modified Green's function and
  Coulomb distorted ingoing wave. By selecting a specific Coulomb
  Green's function, intimately connected with the K-matrix, we have been
  able to devise a calculational scheme that is free from the Coulomb
  singularities. Another advantage of the proposed method is that it
  provides for a simple solution of the Lippmann-Schwinger equation.
  This is a consequence of the fact that for negative energies the
  Coulomb Green's function has a separable representation of the form
  of a Sturmian expansion so that the kernel of the integral equation
  becomes degenerate. The task of solving the Lippmann-Schwinger
  equation is thereby reduced to that of solving a system of linear algebraic
  equations. The  method can be applied either for calculating
  the scattering matrix, or the K-matrix below threshold.
  In particular, this
  approach can be used for locating the poles of the
  scattering matrix what is equivalent to solving the bound state
  problem. The K-matrix yields the effective range function and
  allows to extract the low-energy scattering parameters.
  \par
  The plan of the presentation is as follows. In the next section
  we outline the Lippmann-Schwinger equation formalism necessary for
  calculating the scattering matrix as well as  the K-matrix.
  In section 3, we employ the Sturmian expansion of the Coulomb
  Green's function for converting the underlying integral equations
  into a system of linear algebraic equations. Finally, in the last
  section we discuss possible applications and examine numerically
  the performance of the presented method.
 \section{Effective range function}
 Consider a two-body, two-potential quantum mechanical
 scattering problem with   spherical symmetry where the interaction is 
 a superposition of the Coulomb potential plus a short ranged nuclear
 potential $V(r)$. It will be assumed in the following that the
 nuclear potential for large separations $r$ 
 has at least exponential fall-off $\sim \rme^{-\lambda\,r}$ 
 with $\lambda>0$ being the inverse range parameter of
 the nuclear force. For small $r$, we assume that the behaviour of
 $V(r)$ is not worse than $r^{-2}$.
 The Coulomb potential
 corresponding to point-like charges is taken as $\alpha\, Z/r$
 where $Z$ is dimensionless strength parameter $Z<0$($Z>0$) for attraction
 (repulsion) and $\alpha$ is the fine structure constant.
 The effective range function is defined as the
 inverse of the K-matrix $K_{\ell}(k)$
 and is related to the scattering matrix by
 means of a generalized Heitler formula
 \begin{equation}
 \label{e1}
 \frac{1}{K_{\ell}(k)}= \frac{1}{\tau_{\ell}(k)}+g_{\ell}(k)
 \end{equation}
 where $\ell$ denotes the orbital momentum
 and the independent complex variable $k$ on the positive real axis becomes  
 the physical c.m. momentum and the complex function $g_{\ell}(k)$ can
 be interpreted as a generalized barrier penetration factor. This function
 has been studied in several papers
 \cite{Hull,Hamil,Cornil,Humblet,Lambert,Kok}, and,
 $g_{\ell}(k)=\rmi \, k^{2\ell+1}/[(2\ell+1)!!]^{2}$ for $Z=0$, whereas 
 for $Z\neq 0$ its explicit form is
 \begin{eqnarray}
 \label{e2}
 \fl
 g_{\ell}(k)=(2\mu\alpha Z)^{2\ell+1}/[(2\ell+1)!]^2
 \prod_{m=0}^{\ell}(1+m^{2}/\eta^{2}) 
 \nonumber \\ \lo{\times}
 \left \{\Half\left [\psi(1+\rmi  \eta)+\psi(1-\rmi  \eta) \right ]
 -(\pi/2)\cot(\rmi \eta\pi)-\log(\rmi \mu |Z|/k) \right \},
 \end{eqnarray}
 where $\eta=\mu \alpha Z/k$,  $\mu$ is the reduced mass of the
 system and $\psi$ denotes the logarithmic derivative of the gamma function. 
 We choose the units system where $\hbar = c =1$.  
 In \Ref{e1} the complex function $\tau_{\ell}(k)$
 is the scattering matrix that will be obtained from the solution
 of the Lippmann-Schwinger equation incorporating the 
 appropriate boundary condition
 \begin{equation}
 \label{e3}
 u_{\ell}(k,r)=\phi(k,r)+\int_{0}^{\infty}\Green{+}
  \, V(r')\, u_{\ell}(k,r')\, \rmd r',
 \end{equation}
 where $u_{\ell}(k,r)$ is the sought for wave function. In order
 to account for the Coulomb interaction, in \Ref{e3} the ingoing
 plane wave has been replaced by the regular Coulomb wave function 
 $\phi_{\ell}(k,r)$ and, respectively, the free resolvent by 
 the Coulomb Green's function $\Green{+}$ 
 satisfying the outgoing wave boundary condition.
 It is easily checked that acting on both sides of \Ref{e3} with the
 Coulomb Hamiltonian, we retrieve the Schr\"{o}dinger equation 
 appropriate for the two-potential problem under consideration.
 \par
 Given the solution of \Ref{e3}, the scattering
 matrix $\tau_{\ell}(k)$ is obtained from the formula
 \begin{equation}
 \label{e5}
 \tau_{\ell}(k)=-2\mu\,\int_{0}^{\infty}\phi_{\ell}(k,r)
 \,V(r)\,u_{\ell}(k,r)\,\rmd r,
 \end{equation}
 where the regular Coulomb solution $\phi_{\ell}(k,r)$ 
 satisfying the following boundary condition at $r=0$ 
 \[
 \lim_{r \rightarrow 0}\,r^{-\ell -1}\phi_{\ell}(k,r)=1,
 \]
 is well known \cite{Lambert} and can be expressed analytically as
 \begin{equation} 
 \phi_{\ell}(k,r)=r^{\ell +1} \rme^{\pm \rmi   kr } \mbox{}_{1}F_{1}
 (\ell +1 \pm \rmi   \eta,2\ell +2; \mp 2 \rmi  kr ),
 \label{e4}
 \end{equation}
 where $ \mbox{}_{1}F_{1}(a,b;z)$ denotes the confluent
 hypergeometric function \cite{Abramowitz}.
 Since the boundary condition is independent
 of $k$, it is evident that   $\phi_{\ell}(k,r)$ must be 
 entire analytic function of $k^{2}$ 
 for every non-negative value of $r$ and is real for real $k^{2}$.
 \par
 For physical momenta $k$ the function $\tau_{\ell}(k)$ is related to
 the Coulomb modified nuclear phase shift $\delta_{\ell}$
 \begin{equation}
 \label{e6}
 1/\tau_{\ell}(k)= C_{\ell}(\eta)^{2}\,k^{2\ell}\,
 (k\,\cot{\delta_{\ell}}-\rmi \, k) 
 \end{equation}
 where
 \begin{equation}  
  C_{\ell}(\eta)^{2}\,k^{2\ell}= C_{0}(\eta)^{2}\;
  \frac{(2\mu\alpha Z)^{2\ell+1}}{[(2\ell+1)!]^{2}}
  \prod_{m=0}^{\ell}\left( 1+\frac{m^{2}}{\eta^{2}}\right ),
 \label{e6a}
 \end{equation}
 and
 \begin{equation}
 C_{0}(\eta)^{2}=2\pi\eta/\left [ \exp(2\pi\eta)-1 \right ].
 \label{e6b}
 \end{equation}
 Although for physical momenta $k$  the functions $1/\tau_{\ell}(k)$ 
 and  $g_{\ell}(k)$ are both complex but their imaginary parts are 
 equal in magnitude and
 have opposite signs. Indeed, in this case from \Ref{e2}, one has
 \begin{equation}
 \label{e7}
 g_{\ell}(k)= C_{\ell}(\eta)^{2}\,k^{2\ell}\,
 \left\{ 2\mu\alpha Z [{\rm Re}\,\psi(1+\rmi \eta)-\log(|\eta|)]/
 C_{0}(\eta)^{2} +\rmi  k \right \}
 \end{equation}
 so that using \Ref{e6} and \Ref{e7} in \Ref{e1}, 
 the imaginary parts  are cancelled and the resulting
 effective range function \Ref{e1} is real, as it should be. However, the
 effective range function is known to be real analytic function of $k^{2}$,
 so that it must be real also on the imaginary axis. For $k$ located on
 the positive imaginary axis the Green's function and  $\phi_{\ell}(k,r)$
 are both real what implies that $\tau_{\ell}(k)$ is necessarily real
 as well. Also the  function $g_{\ell}(k)$ is real on the imaginary
 axis 
 \begin{eqnarray}
 \label{e8}
 \fl
 g_{\ell}(k)=(2\mu\alpha Z)^{2\ell+1}/[(2\ell+1)!]^2
 \prod_{m=0}^{\ell}(1-m^{2}/\bar{\eta}^{2})
 \nonumber \\ \lo{\times}
 \left \{\Half\left [\psi(1+\bar{\eta})+\psi(1-\bar{\eta}) \right ]
 -(\pi/2)\cot(\bar{\eta}\pi)-\log(|\bar{\eta}|)\right \},
 \end{eqnarray}
 with $\bar{\eta}=\mu \alpha Z/|k|$ and this ensures that the 
 ensuing effective range function is real. It should be noted
 that in \Ref{e8} the poles
 in $\psi(1-\bar{\eta})$ are cancelled with those from the cotangent term
 and only the poles  coming from $\psi(1+\bar{\eta})$ will survive, they
 correspond to the Coulomb bound states.
 \par
 To complete the above scheme for calculating the effective range
 function we need the explicit formula for the Coulomb Green's function.
 The latter involves a product of the regular $F_{\ell}(\eta,\rho)$ and the
 irregular $G_{\ell}(\eta,\rho)$
 Coulomb wave functions (with $\rho=kr$) defined in \cite{Abramowitz},
 and takes the form
 \begin{equation}
 \label{e8a}
  \Green{+} = -(2\mu/k)
 \left [G_{\ell}(\eta,\rho_{>})+\rmi \;F_{\ell}(\eta,\rho_{>})\right ]
  \; F_{\ell}(\eta,\rho_{<}),
 \end{equation}
 where $ F_{\ell}(\eta,\rho)=C_{\ell}(\eta)\,k^{\ell+1}\phi_{\ell}(k,r)$.
 \par
 Actually, the K-matrix can be calculated directly, i.e. without
 reference to the scattering matrix $\tau_{\ell}(k)$. To this end,
 a different Green's function that is real analytic function
 of $k^{2}$ will be introduced.
  Such Green's function, denoted hereafter
 as $G^{K}_{\ell}$
 to emphasize its connection with the K-matrix, can be easily
 constructed  and reads
 \begin{equation}
 \label{e9}
 \Green{K} = \Green{+}
  +2\mu\,g_{\ell}(k)\,\phi_{\ell}(k,r)\,\phi_{\ell}(k,r'). 
 \end{equation}
 Indeed, the above Green's function has all what it takes. 
 For physical momenta the imaginary parts in the two terms on the
 right hand side of \Ref{e9} cancel each other and 
 $\Green{K}$ is real. For $k$ values located on the
 positive imaginary axis both terms are real and so is
 $\Green{K}$. Using spectral representation
 of $\Green{+}$, it is easy to verify that
 the Coulomb bound state poles, present in $\Green{+} $ 
 for attractive Coulomb potential, will be exactly cancelled with
 the corresponding poles that occur in $g_{\ell}(k)$ (cf.  \Ref{e8}).
 As a result, the Green's function $\Green{K} $
 given by \Ref{e9} is free from the Coulomb singularities and
 is a real analytic function of $k^{2}$ in any finite
 part of the complex k-plane for non-negative $r$ and $r'$. The K-matrix
 will be obtained from the solution of a  Lippmann-Schwinger equation that
 parallels equation \Ref{e3}
 \begin{equation}
 \label{e10}
 w_{\ell}(k,r)=\phi(k,r)+\int_{0}^{\infty}\Green{K} 
  \, V(r')\, w_{\ell}(k,r')\, \rmd r',
 \end{equation}
 where $w_{\ell}(k,r)$ is a new wave function
 that differs from $u_{\ell}(k,r)$ by a constant factor. Given the solution
 of \Ref{e10}, the K-matrix is calculated from the formula
 \begin{equation}
 \label{e11}
 K_{\ell}(k)=-2\mu\,\int_{0}^{\infty}\phi_{\ell}(k,r)
 \,V(r)\,w_{\ell}(k,r)\,\rmd r,
 \end{equation}
 and the above function is meromorphic in any finite part of
 the k-plane. It is easy to check that setting 
 $w_{\ell}(k,r)=[1-g_{\ell}(k)\,K_{\ell}(k)]\,u_{\ell}(k,r)$ in
 \Ref{e10} and using \Ref{e3} -- \Ref{e4}, we immediately retrieve
 the Heitler formula \Ref{e1}. This shows that the Green's function
 \Ref{e9} is unique for a given choice of $g_{\ell}(k)$. It should
 be recalled here that $g_{\ell}(k)$ is not free from ambiguities
 and can be determined only up to a polynomial in $k^{2}$ 
 (cf. \cite{Hamil, Cornil}) but
 the latter is normally set equal to zero. This has been also our choice
 in the present work.
 \par
 The Green's function \Ref{e9} can be also cast into an alternative
 form which is on a par with \Ref{e8a}
 \begin{equation}
 \label{e12}
 \Green{K} = 
  -2\mu\,\phi_{\ell}(k,r_{<})\,\theta_{\ell}(k,r_{>}), 
 \end{equation}
 where the imposed asymptotic boundary condition has been made explicit. 
 The irregular Coulomb wave function $\theta_{\ell}(k,r)$ present in
 \Ref{e12} has been
 introduced in \cite{Humblet,Lambert}. The
 latter is a real analytic function of $k^{2}$ for non-negative $r$
 and can  be expressed as a superposition of the standard
 Coulomb functions
 \[
 \theta_{\ell}(k,r)=C_{\ell}(\eta)\,k^{\ell}\,G_{\ell}(\eta,\rho)
 +{\rm Re}g_{\ell}(k)\,\phi_{\ell}(k,r).
 \]
 Admittedly, the asymptotic behaviour ($r \to \infty $)
 of this function is rather complicated
 but it is worth to pay such a price in view of the fact that   
 the Green's function given by \Ref{e12} is manifestly a real analytic
 function of $k^{2}$, and, as such, can be systematically expanded in
 a Taylor series in powers of $k^{2}$. The form \Ref{e12} is particularly
 useful for calculating the K-matrix at $k=0$. Following \cite{Lambert},
 one has
 \begin{equation}
 \label{e13}
 \fl
 \phi_{\ell}(0,r)=
 \cases{
 [(2\ell+1)!/\beta^{\ell+1}]\sqrt{\beta\,r}
		  \;   I_{2\ell+1}(2\sqrt{\beta\,r})  & for $ Z>0$\\
  r^{\ell+1}                                          & for $ Z=0$\\
  [(2\ell+1)!/\beta^{\ell+1}]\sqrt{\beta\,r}
		  \;   J_{2\ell+1}(2\sqrt{\beta\,r})  &for $ Z<0$ \\
 } 
 \end{equation}
 where $\beta=2\mu \alpha |Z|$, and $(J_{2\ell+1},I_{2\ell+1})$ are
 Bessel and modified Bessel functions, respectively \cite{Abramowitz}.
 The function $\theta_{\ell}(0,r)$
 is given in terms of the Neuman $Y_{2\ell+1}$ and modified
 Neuman function $K_{2\ell+1}$, respectively \cite{Abramowitz}
 \begin{equation}
 \label{e15}
 \fl
 \theta_{\ell}(0,r)=
 \cases{
 [\beta^{\ell}/(2\ell+1)!\sqrt{\beta\,r}
  \;  K_{2\ell+1}(2\sqrt{\beta\,r})  & for $ Z>0$  \\
  r^{-\ell}/(2\ell+1)  & for $ Z=0$ \\
 -\pi/2\,[\beta^{\ell}/(2\ell+1)!]\sqrt{\beta\,r}
  \;  Y_{2\ell+1}(2\sqrt{\beta\,r})  & for $ Z<0$
 }
 \end{equation}
 \par
 Since the K-matrix is a real analytic function of $k^{2}$ the effective
 range function can be expanded in a Taylor series in powers of $k^{2}$
 around threshold
 \begin{equation}
 \label{e20}
 \frac{1}{K_{\ell}(k)}=\pm \frac{1}{a_{\ell}}+\Half\,r_{\ell}\,
 k^{2}+\cdots,
 \end{equation}
 where the expansion coefficients for $\ell=0$ have their standard
 meaning, i.e. $a_{0}$ is the scattering length and $r_{0}$ is the
 effective range and the higher order terms coefficients, that have
 been suppressed in \Ref{e20}, are known as shape parameters. Finally,
 the sign ambiguity in \Ref{e20}, reflecting the coexistence in the
 literature of two different definitions of the scattering length,
 has been resolved by adhering to the generally accepted  
 convention that  $(-)$ sign applies for baryon--baryon
 scattering and  $(+)$ sign in all remaining cases.
 \par
 When the scattering matrix has a bound state pole at $k=k_{B}$, then
 from \Ref{e1} we obtain an equation for the binding energy 
 $B=k_{B}^{2}/2\mu$ involving the effective range function continued
 below threshold
 \begin{equation}
 \label{e21}
 \frac{1}{K_{\ell}(k_{B})}=g_{\ell}(k_{B}).
 \end{equation}
 For shallow bound states   the effective range function
 can be continued below threshold by employing the expansion \Ref{e20}.
 Formula \Ref{e21} correlates then experimentally measurable quantities:
 the low-energy scattering parameters and the binding energy in a model
 independent way.
 \section{Sturmian expansion method}
 Following the scheme devised in the preceding section, the scattering
 matrix and the K-matrix will be obtained by solving a dedicated 
 Lippmann-Schwinger equation. Actually, the solution of the latter
 below threshold presents a quite simple task thanks to the
 Sturmian representation of the Coulomb Green's function.
 \par
 The Sturmians, or Sturm-Liouville functions
 \cite{Weinberg,Rotenberg,Chen}
 $S_{\ell \nu}( b ,r)$, where $b$ is a free parameter
 (${\rm Re}\,b>0$) are enumerated by the
 nodal quantum number $\nu=0,1,2,...$ and result as solution of the
 differential equation
 \begin{equation}
 \left [\frac{\rmd^{2}}{\rmd r^{2}}-\frac{\ell (\ell +1)}{r^{2}}+
 \frac{2 b (\nu+ \ell +1)}{r}- b ^{2}\right ]S_{\ell \nu}( b ,r)=0,
 \label{s11}
 \end{equation}
 with the following boundary conditions 
 imposed, respectively, at zero, and at infinity
 \begin{eqnarray}
 \lim_{r \rightarrow 0}S_{\ell \nu}( b ,r) & \sim &  r^{\ell +1};
 \label{s13}
 \\
 \lim_{r \rightarrow \infty}S_{\ell \nu}( b ,r) & \sim &  \rme^{- b  r}.
 \label{s14}
 \end{eqnarray}
 The eigenfunctions are given in a closed form
 \begin{equation}
 S_{\ell \nu}( b ,r)=\left [\nu!/(2 \ell +1+\nu)!\right ]^{1/2}
 (2 b  r)^{\ell +1} \rme^{- b  r}L_{\nu}^{2 \ell +1}(2 b  r),
 \label{s15}
 \end{equation}
 where $L_{\nu}^{2 \ell +1}$ denotes 
 Laguerre  polynomial. The functions \Ref{s15}
 are orthogonal and form a complete set with respect to the weight
 function $1/r$.
 \par
 For $Im \,k >0$ the Coulomb Green's  function $G^{+}_{\ell}(k)$ 
 turns out to be diagonal in a very specific Sturmian basis, namely when
 the parameter $b$ has been set equal to $-\rmi  k$.
 Therefore, in coordinate representation, following \cite{Chen}, we
 have
 \begin{equation}
 \Green{+} =
 \frac{\mu}{\rmi   k} \sum_{\nu=0}^{\infty}
 \, \frac{S_{\ell \nu}(-\rmi   k,r)\, \, S_{\ell \nu}(-\rmi   k,r^{\prime})}
 {\nu+\ell +1+\rmi   \eta}
 \label{s12}
 \end{equation}
 but it should be clearly stated that for real $k$ the above
 series {\em diverges} so that \Ref{s12} is invalid for physical
 momenta.
 Formula \Ref{s12} is quite remarkable in that the Green's function is
 given as a sum of the poles at the Coulomb
 bound states yet the same expression holds for Coulomb repulsion
 as well as in the Coulomb-free case. Unlike the function $g_{\ell}(k)$,
 the Green's function is analytic function of the Coulomb strength
 parameter $Z$.  Most importantly, however,
 the residues in \Ref{s12} factorize what results in a separable
 representation very useful in practical applications.
 In particular, this implies that the kernel of the Lippmann-Schwinger
 equation either \Ref{e3}, or \Ref{e10} becomes degenerate. With such a
 simplification the solution of equation \Ref{e10} is straightforward.
 Setting $k=\rmi \,p$ with real $p>0$ and
 inserting \Ref{s12} together with \Ref{e9} in \Ref{e10}, 
 the solution of \Ref{e10} is
 \begin{equation}
 \label{s1}
 \fl
 w_{\ell}(k,r)=\left [1-g_{\ell}(k)\,K_{\ell}(k)\right ]\phi_{\ell}(k,r)
  -\frac{\mu}{p}\,\sum_{\nu=0}^{\infty}
 \frac{S_{\ell\nu}(p,r)\,X_{\ell\nu}(p)}{\ell+1+\nu+\bar{\eta}},
 \end{equation}
 where 
 \begin{equation}
 \label{s2}
 X_{\ell\nu}(p)=\int_{0}^{\infty}S_{\ell\nu}(p,r)\,V(r)\,
 w_{\ell}(k,r)\,\rmd r.
 \end{equation}
 In order to determine the unknown coefficients $X_{\ell\nu}(p)$,
 as well as the K-matrix,
 the expression \Ref{s1} for $w_{\ell}(k,r)$ is inserted in \Ref{s2}
 and in \Ref{e11}.
 This gives an infinite system of linear algebraic equations
 where the unknowns are the coefficients $X_{\ell\nu}$ together with
 $K_{\ell}(k)$
 \begin{equation}
 \label{s3}
 \fl
 \sum_{\nu^{\prime}=0}^{\infty}\left [\delta_{\nu\nu^{\prime}}+
 \frac{\mu}{p}\frac{\bra \ell\nu|V|\ell\nu^{\prime}\ket}
 {\ell+1+\nu^{\prime}+\bar{\eta}}\right ]\,X_{\ell\nu^{\prime}}(p)
 +g_{\ell}(k)\,A_{\ell\nu}(p)\,K_{\ell}(k)=
 A_{\ell\nu}(p),
 \end{equation}
 where $\nu=0,1,2,...$, and this 
 infinite set is supplemented by the equation 
 \begin{equation}
 \label{s4}
 \fl
 \frac{\mu}{p}\,\sum_{\nu^{\prime}=0}^{\infty}
 \frac{A_{\ell\nu^{\prime}}(p)\,X_{\ell\nu^{\prime}}(p)}
 {\ell+1+\nu^{\prime}+\bar{\eta}}
 +\left [ g_{\ell}(k)\,\bra \phi_{\ell}|V| \phi_{\ell}\ket
 -(1/2\mu) \right ]\,K_{\ell}(k)=
 \bra \phi_{\ell}|V| \phi_{\ell}\ket.
 \end{equation}
 In \Ref{s3} and \Ref{s4} we have introduced the following
 abbreviations
 \begin{equation}
 \label{s5}
 A_{\ell\nu}(p)=\int_{0}^{\infty}S_{\ell\nu}(p,r)\,V(r)\,
 \phi_{\ell}(k,r)\,\rmd r,
 \end{equation}
  $\bra \ell\nu|V|\ell\nu^{\prime}\ket$ denotes the expectation
 value of the potential in the Sturmian basis
 \begin{equation}
 \label{s6}
 \bra \ell\nu|V|\ell\nu^{\prime}\ket =
 \int_{0}^{\infty}S_{\ell\nu}(p,r)\,V(r)\, 
 S_{\ell\nu^{\prime}} (p,r)\,\rmd r,
 \end{equation}
 and $\bra \phi_{\ell}|V| \phi_{\ell}\ket$ is the Born term 
 \begin{equation}
 \label{s7}
 \bra \phi_{\ell}|V| \phi_{\ell}\ket=
 \int_{0}^{\infty}\phi_{\ell}(k,r)\,V(r)\, 
 \phi_{\ell}(k,r)\,\rmd r.
 \end{equation}
 The integrals \Ref{s5} -- \Ref{s7},
 carrying important information about the potential, 
 should be regarded as input and
 to proceed further it is sufficient to 
 truncate the infinite Sturmian basis 
 approximating it by a finite dimensional one. When the resulting finite
 set of linear algebraic equations has been solved by any of the standard
 methods, one of the solutions contains the K-matrix.
 \par
 The scattering matrix $\tau_{\ell}(k)$ can be calculated in a similar
 way. By inserting \Ref{s12} in \Ref{e3}, we get
 \begin{equation}
 \label{s8}
 u_{\ell}(k,r)=\phi_{\ell}(k,r)
  -\frac{\mu}{p}\,\sum_{\nu=0}^{\infty}
 \frac{S_{\ell\nu}(p,r)\,Y_{\ell\nu}(p)}{\ell+1+\nu+\bar{\eta}},
 \end{equation}
 where the new expansion coefficients defined as
 \begin{equation}
 \label{s9}
 Y_{\ell\nu}(p)=\int_{0}^{\infty}S_{\ell\nu}(p,r)\,V(r)\,
 u_{\ell}(k,r)\,\rmd r.
 \end{equation}
 are obtained by solving the following system of linear algebraic
 equations
 \begin{equation}
 \label{s10}
 \sum_{\nu^{\prime}=0}^{\infty}\left [\delta_{\nu\nu^{\prime}}+
 \frac{\mu}{p}\frac{\bra \ell\nu|V|\ell\nu^{\prime}\ket}
 {\ell+1+\nu^{\prime}+\bar{\eta}}\right ]\,Y_{\ell\nu^{\prime}}(p)
  =A_{\ell\nu}(p).
 \end{equation}
 Using \Ref{s8} in \Ref{e4}, we arrive at  the ultimate expression for
 the scattering matrix
 \begin{equation}
 \label{s16}
 \tau_{\ell}(k)=-2\mu\left [ \bra \phi_{\ell}|V| \phi_{\ell}\ket-
 \frac{\mu}{p}\,\sum_{\nu^{\prime}=0}^{\infty}
 \frac{A_{\ell\nu^{\prime}}(p)\,Y_{\ell\nu^{\prime}}(p)}
 {\ell+1+\nu^{\prime}+\bar{\eta}} \right ].
 \end{equation}
 Similarly as before, for practical purposes the infinite Sturmian basis
 must be approximated by a finite dimensional one.
 The values of $k$ for which the determinant of the system \Ref{s10}
 vanishes correspond to the bound states. For attractive Coulomb potential
 ($Z<0$) they are infinite in number, all shifted from the pure Coulomb
 position owing to the nuclear interaction. For $Z \ge 0$ the bound
 states may, or may not exist, depending upon the nature of the
 nuclear potential.
 Since the case of bound states has been discussed
 elsewhere \cite{DL}, we assume in the following that for the considered 
 $k$ value the system of equations \Ref{s10} is non-singular. 
 Once $\tau_{\ell}(k)$ has been determined, the effective range function
 may be obtained by making use of Heitler formula but this approach
 is equivalent to solving the system of equations \Ref{s3} and \Ref{s4}.
 The determinant of that system may also vanish at some values of $k$
 and they correspond to poles of the K-matrix. Clearly, the poles of the
 scattering matrix are distinct from the poles of the K-matrix.
 \par
 The proposed scheme has, of course, its limitations and breaks down
 not only for large $p$ but also for very small $p$. The large $p$ limit is
 rather obvious and stems from the fact that for large $r$ the regular
 Coulomb wave function $\phi_{\ell}(k,r)$ increases like $\sim \rme^{pr}$
 and therefore the Born term integral \Ref{s7} diverges for
 $p>\lambda/2$. Consequently, the largest admitted $p$ value is determined
 by the range of the nuclear potential. There is also a lower bound 
 for $p$, as in the limit of small $p$ the convergence rate of the
 Sturmian expansion gradually
 deteriorates eventually rendering the whole scheme
 impractical. The lower limit for $p$, according to our experience,
 depends upon the characteristics of the nuclear potential. 
 In this situation, the
 effective range function must be extrapolated to zero. If the
 extrapolation is not satisfactory the K-matrix at $k=0$ can be
 obtained by solving the Lippmann-Schwinger equation with the Green's
 function given by \Ref{e12}. Although
 the kernel is not degenerate in that case but the
 resulting integral equation can be easily solved by standard methods.
 Another alternative is a numerical integration of the appropriate
 Schr\"{o}dinger equation. A~stepwise integration of the latter from
 $r=0$ up to some large $r=R$, well beyond the range of the nuclear
 potential, yields the regular solution $u_{\ell}(0,R)$ and the 
 derivative $u_{\ell}^{\prime}(0,R)$. With these two functions
 in hand, the effective range function is
 \[
 \frac{1}{K_{\ell}(0)}= 
 -\frac{W[u_{\ell}(0,R),\theta_{\ell}(0,R)]}
        {W[u_{\ell}(0,R),\phi_{\ell}(0,R)]},
 \]
 where $W[f,g]=f\,g'-g\,f'$ denotes the Wronskian determinant and
 the Coulomb wave functions are given by \Ref{e13} and  \Ref{e15},
 respectively.
 \section{Applications}
 In order to examine the performance of the calculational scheme 
 presented in the preceding section, the
 Sturmian expansion method will be subjected to a rather demanding test.
 To achieve this goal, for the nuclear potential $V(r)$
 we have selected Reid nucleon-nucleon $\mbox{}^{1}S_{0}$ potential
 \cite{Reid}.
 This potential is given as a superposition of three Yukawa potentials,
 has a one-pion-exchange tail and a short range very strong repulsive
 term. Since the shape of this potential shows a good deal of structure
 comprising of attraction and repulsion, the rank $N$ of the Sturmian
 approximation must be large enough to account for the
 rapid variation of the potential. In addition to that, the other
 unfavourable circumstance is that the long one-pion-exchange tail
 greatly reduces the range of the admissible momenta which must not
 exceed $65$~MeV. In the actual computations, we have been able
 to determine the effective range function in the interval
 $20\,{\rm MeV}<p<65\,{\rm MeV}$. The rank of the Sturmian approximation $N$ 
 at which convergence has been attained
 varied from about hundred to several hundred at the low-energy end.
 It is perhaps worth noting that when $V(r)$ has a Yukawa shape
 then all the necessary integrals
 are given in an analytic form. To be more explicit, $A_{\ell\nu}$ is
 obtained from
 \begin{eqnarray}
 \fl
 \int_{0}^{\infty}S_{\ell\nu}(b,r)\, \frac{\rme^{-\lambda\,r}}{r}\,
 \phi_{\ell}(k,r)\,\rmd r=
 \sqrt{\frac{(2\ell+1+\nu)!}{\nu !}}\left [\frac{2\,b}
 {(\lambda+b)^{2}+k^{2}}\right ]^{\ell+1} 
 \left (\frac{\lambda-b+\rmi \,k}{\lambda+b+\rmi \,k}\right )^{\nu}
  \nonumber \\
 \lo{\times}
 \left (\frac{\lambda+b-\rmi \,k}{\lambda+b+\rmi \,k}\right )^
 {\rmi \eta}\;\mbox{}_2F_{1}(-\nu,\ell+1-\rmi \eta;2\ell+2;1-z) 
 \label{R1}
 \end{eqnarray}
 where $z=[\lambda^{2}-(b+\rmi  k)^{2}]/ [\lambda^{2}-(b-\rmi  k)^{2}]$.
 The expectation value of the potential in the Sturmian basis is obtained from 
 \begin{eqnarray} \fl
 \int_{0}^{\infty}S_{\ell\nu}(b,r)\, \frac{\rme^{-\lambda\,r}}{r} \,
 S_{\ell\nu^{\prime}} (b,r)\,\rmd r=
 \frac{1}{(2\ell+1)!}
 \sqrt{\frac{(2\ell+1+\nu)!\;(2\ell+1+\nu')!}{\nu!\;\nu'!}}
  \nonumber \\
 \lo{\times} x^{2\ell+2}\,(x+1)^{-2\ell-2-\nu-\nu'}\;
 \mbox{}_{2}F_{1}(-\nu,-\nu';2\ell+2;x^{2}),
 \label{R2}
 \end{eqnarray}
 where $x=2b/\lambda$. Finally, the Born term, takes the form 
 \begin{eqnarray}  
 \int_{0}^{\infty}\phi_{\ell}(k,r)\,
 \frac{\rme^{-\lambda\,r}}{r} \, \phi_{\ell}(k,r)\,\rmd r=
 \frac{(2\ell+1)!}{(\lambda^{2}+4k^{2})^{\ell+1}}
 \left (\frac{\lambda-2\rmi  k}{\lambda+2\rmi  k}\right )^{\rmi \eta}
  \nonumber \\  \hspace{1cm} \times \;
 \mbox{}_{2}F_{1}(\ell+1-\rmi \eta,\ell+1+\rmi \eta; 2\ell+2;y),
 \label{R3}
 \end{eqnarray}
 where $ y=4k^{2}/(\lambda^{2}+4k^{2})$. In the above expressions
 we assumed that $\lambda^{2}>4k^{2}$ and
 $\mbox{}_{2}F_{1}(a,b;c;z)$ denotes the hypergeometric function.
 \par
 The effective range functions resulting from our computations are
 presented in Figure~1  as functions of the c.m. energy $E=k^{2}/2\mu$.
 The calculations have been carried out for three values of the 
 Coulomb strength parameter $Z$: $Z=\pm 1$ and $Z=0$.
 Strictly speaking, attractive
 Coulomb potential would be appropriate for the proton-antiproton pair.
 However, in that case the Reid potential, devised to fit the nucleon-nucleon
 scattering data, is not applicable
 and the inferred effective range parameters have no relevance
 to proton-antiproton scattering. Nevertheless, this case
 is of theoretical interest and has been included here for testing purposes.
 In Figure~1 the continuous curves above threshold result from a numerical
 integration of the appropriate Schr\"{o}dinger equation. The entries below
 threshold, depicted by asterisk symbols on the plot, have been obtained
 by using the Sturmian expansion method. As expected, close to $E=0$
 the values of the effective range function  calculated below and above
 threshold appear to be lying on the same straight line. 
 In order to qualify that statement,
 the values calculated below threshold have been subsequently used
 to extrapolate each of the three functions to zero. This gave  the 
 intersection points at zero together with the slopes of the curves. 
 The scattering lengths and effective ranges, extracted that way
 with the aid of \Ref{e20},
 are presented in Table~I (referred to as  ''below'').
 For comparison, we give also their counterparts  
 calculated from the above threshold solutions of the
 Sch\"{o}dinger equation (referred to as "above" in Table~I).
 The agreement between the below threshold and
 the above threshold computations is in all three cases very good.
 The error on $r_{0}$ is slightly bigger than that on $a_{0}$,
 as the computation of the former involves the differences of the 
 function values what results in some loss of accuracy.
 The effective range seems to be little affected by the Coulomb
 interaction what is also reflected in Figure~1 in that the three curves
 have almost the same slope near zero. By contrast, the Coulomb 
 corrections to the scattering length are quite large especially
 for Coulomb attraction when $a_{0}$ has been pushed far away
 to positive values.
 The K-matrix, unlike the effective range function, shows rapid
 variation as function of energy and in all three cases exhibits
 a pole not far from the origin. At the pole the effective range
 function goes to zero, so that the actual positions of the poles
 of the K-matrix are apparent from Figure~1.
 It is interesting to note that for attractive Coulomb potential
 the pole of the K-matrix lies in the physical region, very close to
 threshold. This example is rather instructive, because it shows that
 any phenomenological attempt to approximate here the K-matrix by
 a polynomial in $k^{2}$, with coefficients to be fitted to the
 low-energy data, would have a disastrous effect. The problem is
 immediately alleviated when instead of the K-matrix the effective
 range function is employed.
 \par
 The two-potential problem considered in this work can be easily
 generalized to comprise complex nuclear potentials.
 Such potentials appear in various problems of nuclear physics
 and, in particular, they have been widely used for a phenomenological 
 description of the hadronic atoms phenomena.
 The introduction of a complex nuclear potential into the description
 of a hadronic atom has two observable effects: a shift $\epsilon$ in the
 energy of the otherwise hydrogen-like level position, and a
 broadening  $\Gamma$ of the atomic level arising from the fact that
 the system can now decay through the absorptive strong interaction.
 When the orbital momentum is large enough the level shift in a hadronic
 atom is usually quite small in comparison with the Coulomb energy. This
 observation has prompted several authors 
   (cf. \cite{Deser, True, Parten, Lambert, AD2, Mand}) 
 to introduce small shift
 approximation (SSA) based on the idea that if atomic bound state pole has
 been shifted from the Coulomb position $k_{n}=-\rmi \,\mu\alpha\, Z/n$,
 where $n$ is the principal quantum number, by $\delta k$ then
 $|\delta k/k_{n}|$ can be regarded as a small expansion parameter.
 One of the implications of SSA is the shift formula expressing the energy
 shift in a hadronic atom in terms of hadron-nucleus low-energy
 scattering parameters. This formula relating only experimentally
 measurable quantities without reference to the underlying nuclear
 potential has very important phenomenological applications.
 \par
 In order to derive  the shift formula we are going to obtain an  
 approximate solution of the bound state equation \Ref{e21}. 
 As mentioned before,
 the function $g_{\ell}(k)$ that enters \Ref{e21} 
 is singular at $k=k_{n}$ and for $k$ such that
 $|k-k_{n}|$ is small in comparison with the separation of the
 unperturbed levels
 $|k_{n}-k_{n+1}|$, we can approximate $g_{\ell}(k)$ by a single
 pole term
 \begin{equation}
 g_{\ell}(k)\approx \frac{k_{n}\,R_{\ell n}}{k-k_{n}}.
 \label{z31}
 \end{equation}
 The residue $R_{\ell n}$ can be worked out from the
 definition of $g_{\ell}(k)$ and reads
 \begin{equation}
 R_{\ell n}=\Big \{2|k_{n}|^{2}\int_{0}^{\infty}\phi_{\ell}(k_{n},r)^{2}
 \,\rmd r \Big \}^{-1}
 \label{z32}
 \end{equation}
 where the normalization integral, is
 \begin{equation}
 \int_{0}^{\infty}\phi_{\ell}(k_{n},r)^{2}\, \rmd r=
 \frac{n(n-\ell-1)!
 \left [(2\ell +1)!!\,\ell !\right ]^{2}}{4|k_{n}|^{2 \ell+3} (\ell +n)!}.
 \label{z21}
 \end{equation}
 In order to obtain an approximate solution of equation \Ref{e21} it is
 assumed that the left hand side
 is a slowly varying function of $k$ so that
 $1/ K_{\ell} (k)$ may be replaced by $ 1/K_{\ell} (k_{n})$.
 Using the single pole approximation \Ref{z31}, the solution of
  \Ref{e21} is
 \begin{equation}
 k\approx k_{n}[1+ K_{\ell} (k_{n})\,R_{\ell n}],
 \label{z33}
 \end{equation}
 and this yields the level shift formula in its most general
 form
 \begin{equation}
 \delta E=-\epsilon-\rmi  \Half\Gamma=E-E_{n}
 \approx - K_{\ell} (k_{n})
 \left\{2\mu\int_{0}^{\infty}\phi_{\ell}(k_{n},r)^{2}\, \rmd r\right\}^{-1}
 \label{z20}
 \end{equation}
 where $E_{n}=k_{n}^{2}/2 \mu$
 and it is understood that for complex $V(r)$ the K-matrix is
 necessarily also complex.
 Approximating in \Ref{z20} the 
 K-matrix by the scattering length  one obtains the
 celebrated Deser--Trueman  formula \cite{Deser, True}.
 However, the latter approximation is hardly necessary as  the K-matrix 
 below threshold can be calculated
 with the same amount of labour by using the Sturmian expansion
 method described in the preceding section. The other alternative,
 that  does not require the SSA ansatz,  
 is to locate the poles of the scattering matrix. Using Sturmian
 expansion, the problem reduces to finding the values of $k$ at which 
 the determinant of the system  \Ref{s10} vanishes (cf. \cite{DL}). 
 \par
 In conclusion, we have presented a simple and efficient method
 for calculating the effective range function below threshold and the same
 calculational scheme can be used for solving the bound state problem.
 The main advantage of the proposed method lies in the exact treatment of the
 Coulomb interaction what is accomplished by using
 the Lippmann-Schwinger equation whose kernel involves the Coulomb
 Green's function. Since this kernel below threshold assumes degenerate
 form, the solution of the integral equation is straightforward.
 The essential point is that the Coulomb Green's function and the
 regular Coulomb wave function, constituting input to the
 Lippmann-Schwinger equation, are real analytic functions of $k^{2}$,
 free from Coulomb singularities. Both are given in analytic form
 suitable for numerical computations and the irregular Coulomb wave
 function is never needed.

  \Tables
 \begin{table}
  \vspace*{5cm}
 \caption
 {Coulomb-modified low-energy scattering parameters
  for   Reid $\mbox{}^{1}S_{0}$  NN potential calculated 
  below and above threshold. All entries are in fm. \label{Tab}}
  \begin{indented}
  \lineup
  \item[]\begin{tabular}{lllllll}
 \br
  \multicolumn{1}{c}{}&
 \multicolumn{2}{c}{$Z=1$}&
 \multicolumn{2}{c}{$Z=0$}&
 \multicolumn{2}{c}{$Z=-1$}\\
 \mr
 \ns
 \multicolumn{1}{l}{method}  &
 \multicolumn{1}{c} {$a_{0}$}&
 \multicolumn{1}{c} {$r_{0}$}&
 \multicolumn{1}{c} {$a_{0}$}&
 \multicolumn{1}{c} {$r_{0}$}&
 \multicolumn{1}{c} {$a_{0}$}&
 \multicolumn{1}{c} {$r_{0}$}\\
 \mr
 below & -7.77 & 2.75 & -17.1 & 2.83 & 148. & 2.91 \\ 
 \mr
 above & -7.77 & 2.69 & -17.1 & 2.78 & 147. & 2.88 \\ 
 \br
 \end{tabular}
           \end{indented}
 \end{table}
  \Figures
  \begin{figure}[ht]
  \centering
  \includegraphics[width=0.8\textwidth]{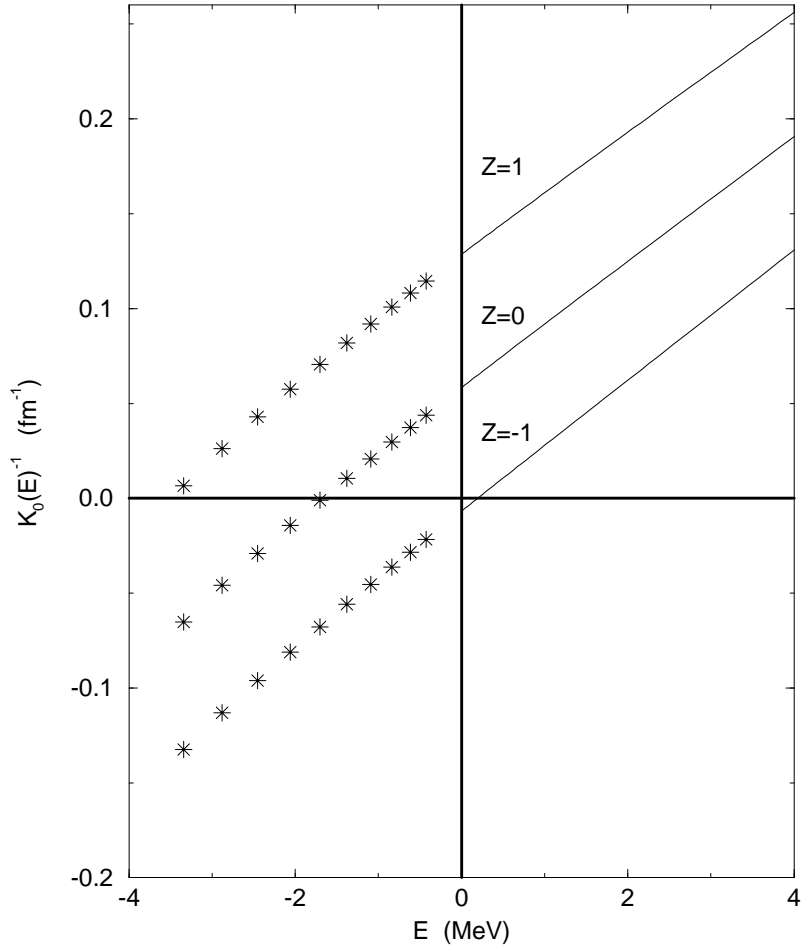}
  \caption[]{ Coulomb-modified effective range function vs. c.m. energy for
   Reid $\mbox{}^{1}S_{0}$ NN  potential. The different curves correspond to  
   Coulomb attraction ($Z=-1$),  Coulomb repulsion  ($Z=1$)
   and  Coulomb-free case ($Z=0$). }
  \label{fig1}
  \end{figure}
   \end{document}